\documentclass[referee]{aa} 

\usepackage{graphicx}
\usepackage{natbib}
\usepackage{ulem}
\usepackage{txfonts}

\usepackage{rotating}
\usepackage{color}
\usepackage{amssymb}
\usepackage{tensor}

\newcommand{\be}{\begin{equation}}
\newcommand{\ee}{\end{equation}}

\def\JGR{J. Geophys. Res.}

\def\PPCF{Plasma Phys. Control. Fusion}

\begin{document}
\authorrunning{Silvia Perri et al.}
\titlerunning{Superdiffusion at heliospheric shocks}




\title{Parameter estimation of superdiffusive motion of energetic particles upstream of heliospheric shocks}


\author{Silvia Perri \inst{1}, Gaetano Zimbardo
\inst{1}
\and
Frederic Effenberger\inst{2}
\and
Horst Fichtner\inst{3}
}

\institute{Department of Physics, University of Calabria, Ponte P. Bucci, Cubo 31C, I-87036, Rende, Italy\\
\email{silvia.perri@fis.unical.it}
\and
Department of Mathematics, University of Waikato, PB3105, Hamilton, New Zealand
\and
Institut f\"ur Theoretische Physik IV, Ruhr-Universit\"at Bochum, 
         Universit\"atsstrasse 150, 44780 Bochum, Germany}

\abstract
{In-situ spacecraft observations recently suggested that the transport of energetic particles 
accelerated at heliospheric shocks can be anomalous, i.e. the mean square displacement can grow non-linearly in time. In particular, a new analysis technique has permitted the study of particle transport properties from energetic particle time profiles upstream
of interplanetary shocks. Indeed, the time/spatial power laws of the differential intensity upstream of several shocks are indicative of superdiffusion.} 
{A complete determination of the key parameters of superdiffusive transport comprises the power-law index, the superdiffusion coefficient, the related transition scale at which the energetic particle profiles turn to decay as power laws, and the energy spectral index of the shock accelerated particles.} 
{Assuming large-scale spatial homogeneity of the background plasma, the power-law behaviour can been derived from both a (microscopic) propagator formalism and a (macroscopic) fractional transport equation. We compare the two approaches and find a relation between the diffusion coefficients used in the two formalisms. Based on
the assumption of superdiffusive transport, we quantitatively derive these parameters by studying energetic particle profiles observed by the Ulysses and Voyager 2 spacecraft upstream of shocks in the heliosphere, for which a superdiffusive particle transport has previously been observed. Further, we have jointly studied the electron energy spectra, comparing the values of the spectral indices observed with those predicted by the standard diffusive shock acceleration theory and by a model based on superdiffusive transport.} 
{For a number of interplanetary shocks and for the solar wind termination shock, for the first time we obtain the anomalous diffusion constants and the scale at which the probability of particle free paths changes to a power-law. The investigation of the particle energy spectra indicates that a shock acceleration theory based on superdiffusive transport better explains observed spectral index values.} 
{This study, together with the analysis of energetic particles upstream of shock waves, allows us to fully determine the transport properties of accelerated particles, even in the case of superdiffusion. This represents a new powerful tool to understand the transport and acceleration processes at astrophysical shocks.} 

\maketitle

\parindent=0.5 cm


\section{Introduction}
The transport of energetic particles in space and astrophysical plasmas is a topic of current interest: for instance, the transport properties determine the measured fluxes of solar energetic particles, related to solar activity, in the geospace environment, as well as the features of particle acceleration at astrophysical shock waves \citep{Duffy95,Kirk96,Ragot97,Perri12a,Zimbardo13,Dalla13}. For many years particle acceleration at shocks has been described in the framework of diffusive shock acceleration (DSA). This theory is based on a first-order Fermi acceleration where particles, with Larmor radii larger than the shock thickness, can cross the shock front many times, thanks to interactions with magnetic irregularities that scatter particles back and forth (diffusive motion) across the shock \citep[e.g.][]{Lee82}. This process allows particles to reach high energies. However, DSA cannot always explain energetic particle observations in space. 

It is well known that beside normal diffusion, characterized by Gaussian statistics and by a mean square displacement that grows linearly in time as $\langle x^2(t) \rangle= 2 D t$, with $D=1/3 \lambda v$ where $\lambda$ denotes the mean free path and $v$ is the particle speed, other transport regimes are also possible. In particular, they are denoted as anomalous regimes where $\langle \Delta x^2 \rangle = 2 {\cal D}_{\alpha} t^{\alpha}$ with $\alpha\neq 1$.
In this case, the anomalous diffusion coefficient ${\cal D}_{\alpha}$ has dimensions $[{\cal D}_{\alpha}]=$(length)$^2/$(time)$^{\alpha}$.
More in detail, slower processes ($\alpha<1$) are called subdiffusive, and faster processes ($2>\alpha>1$) are called superdiffusive (we focus our attention on the latter).
Anomalous regimes are characterized by non-Gaussian statistics like L\'evy statistics, which encompasses probability distributions with power-law tails. A number of theoretical tools to study anomalous transport have been used, including L\'evy flights, L\'evy walks, and fractional transport equations \citep[e.g.][]{Klafter87,Zaslavsky02,Metzler04,del-Castillo04,Webb06,Shalchi09,Litvinenko14,Stern14}. 
Numerical simulations of particle propagation in the presence of magnetic turbulence have indicated that anomalous transport can be found, depending on the particle energy and turbulence properties  \citep{Zim05,Zim06,Pommois07,Shalchi07,Tautz10}.

In a series of papers, \citet{Perri07,Perri08,Perri09a,Perri09b} have shown that anomalous, superdiffusive transport of energetic particles accelerated at interplanetary shocks is indeed possible. Those analyses were based on the observation that at few hours upstream of the shocks the energetic particle time profiles are power laws rather than exponentials, the latter being predicted by DSA. Further, the diffusion coefficient and the pitch angle scattering rate were found to be constant, since the magnetic field fluctuations at the resonant scales upstream of the shocks are in a steady state \citep{Perri10,Perri12b}. Standard arguments based on DSA \citep{Bell78} would require non-constant upstream turbulence levels to explain the observed power-law profiles. The latter allow the determination of the exponent of superdiffusion $\alpha$; however, for a complete description of transport, the anomalous diffusion coefficient ${\cal D}_{\alpha}$ has to be determined, too, as well as the scale of transition to a power-law profile. The approach described in \citet{Zimbardo13} is based on a specific type of continuous time random walk i.e. on L\'evy walks, which are random walk processes with a power-law distribution of free path lengths and a coupling between the free path length and travel time. This coupling is appropriate for the description of particles having a finite velocity. Another approach is based on fractional transport equations, that is on the generalization of the standard diffusion equation to the cases where fractional, non-integer derivatives of order $\mu$ are present (i.e. $\partial_t f=\kappa \partial ^{\mu} f/\partial |x|^{\mu}$). In the usual formulation, fractional diffusion equations correspond to L\'evy flights rather than to L\'evy walks, that is to a model where free path lengths and free path travel times are not coupled \citep{Metzler00}. Therefore some differences in the prediction of the two approaches may exist. However, the use of fractional transport equations can be very useful, because they allow for a simple inclusion of external velocity and force fields, like the shock advection speed, and because powerful and well-tested solution techniques do exist for fractional equations \citep{Metzler00, Stern14, Litvinenko14}. 
On the other hand, the order of differentiation and the diffusion constant $\kappa$ in the fractional diffusion equation are often introduced ``ad hoc'' and thus have a phenomenological character \citep{Metzler98}. Therefore, it is important to find a physically based way to relate $\alpha$ to $\mu$ and ${\cal D}_{\alpha}$ to $\kappa$.
In the analysis carried out below, we show that the determination of the length $\ell_0$ that represents the shortest path length for which a power-law probability distribution holds, is fundamental for characterizing the particle transport. Indeed, the values of ${\cal D}_{\alpha}$ and $\kappa$ are found to depend on $\ell_0$.

In this paper, we show for the first time how $\ell_0$, ${\cal D}_{\alpha}$, and $\kappa$ can be obtained from the analysis of energetic particle profiles upstream of shocks. Both corotating interaction region (CIR) shocks observed by Ulysses and the solar wind termination shock (TS) observed by Voyager~2 are considered.

An important consequence of anomalous diffusion for particle acceleration at shocks is that the energy spectral index $\gamma_{DSA}$ predicted by DSA is modified. In particular it yields larger spectral indices (softer particle spectra) in the case of subdiffusion \citep{Kirk96} and smaller spectral indices (harder particle spectra) in case of superdiffusion \citep{Perri12a,Zimbardo13}. In the anomalous framework the spectral indices depend not only on the compression ratio of the shock $r$ (i.e. the ratio between the plasma densities downstream and upstream), but also on the anomalous diffusion exponent $\alpha$. We carry out a comparison between the observed $\gamma_{obs}$ and that obtained from the theory once $r$ and $\alpha$ are known from the shock data \citep{Perri12a,Zimbardo13}. We find that superdiffusion allows for a better explanation of the observed spectral indices.

In Section~2 we give a description of superdiffusion in terms of L\'evy walks and introduce the main parameters that will be determined by data analysis. In Section~3 we give a description of anomalous transport in terms of fractional differential equations indicating the analogies between the two approaches and the relation between the diffusion coefficients ${\cal D}_{\alpha}$ and $\kappa$.
In Section~4 we present the method to extract from spacecraft data the values of a scale parameter $\ell_0$, the superdiffusion coefficients ${\cal D}_{\alpha}$ and $\kappa$, and the energy spectral index $\gamma_{SSA}$, and we discuss the results of the analysis for a number of interplanetary shocks as well as for the TS, at which energetic particle exhibit superdiffusion. In Section~5 we give the conclusions.

\section{Superdiffusion in terms of L\'evy walks}

The interplanetary shocks studied here are large-scale shocks with a radius of curvature of about 5 AU or more. Since the transport and the acceleration processes occur on scales much smaller than $5$ AU, these shocks are regarded as being planar, and the corresponding shock transition structure is considered to be one dimensional.  Therefore, particles making random walks sample variations of the shock structure only in the direction perpendicular to the shock. To facilitate this analysis, we call the dimension perpendicular to the (planar) shock $x$ and study the transport properties along this coordinate. 


Superdiffusion can be described microscopically via a probabilistic approach that involves the so-called L\'evy random walks \citep[e.g.][]{Klafter87}. 
Here, we summarize the main properties of a L\'evy random walk, referring to \citet{Metzler00}, \citet{Metzler04}, \citet{Zimbardo12}, and \citet{Perrone13} for more details, and to \citet{Zimbardo13} for a recent analytical derivation.

The probability $\psi$ for a random walker to make a free path (or jump) of length $\ell$ (forward or backward) in a time 
$t$ is given by \citep{Shlesinger87,Klafter87,Zumofen93}
\be
\psi (\ell, t) = A \, |\ell|^{-\mu-1} \, \delta (|\ell| - vt),  
\;\;\;\;\;\; |\ell| > \ell_0, 
\;\;\;\;\;\; t > 0
\label{eq:psi}
\ee
where the coupling between the jump length and the travel time, as expressed by the delta function, is essential to ensure the finite (and, in particular, constant) particle velocity. 
With respect to our previous works, we changed the notation from $\mu$ to $\mu+1$. In the above expression, the power-law form only applies for $|\ell| > \ell_0 $, with $\ell_0$ a scale parameter. An otherwise bell-shaped profile ensures that $\psi$ does not diverge for $\ell\to 0$. For small values of the parameter ${\mu}$, ``heavy" tails of $\psi$ are obtained, which correspond to a non-negligible probability for particles to perform very long jumps. It is readily verified that for ${\mu} < 2$ the mean square value of $\ell$, and hence the mean free path, is diverging i.e. $\langle \ell^2 \rangle = \int \ell^2 \, \psi (\ell, t) \, d\ell dt \to \infty$, meaning that the validity of the central limit theorem, which is pivotal in normal transport, breaks down
\citep[e.g.][]{Klafter87}. In particular, for $1<{\mu}<2$ we obtain superdiffusion and the parameter ${\mu}$ is directly related to the anomalous diffusion exponent $\alpha$ via
\be
\label{eq:alpha-mutil}
\alpha = 3 - {\mu} 
\ee
\citep{Geisel85,Klafter87,Zumofen93}.

The probability for a `jumping' particle to be at a position $x$ at time $t$
is given by a propagator $P(x,t)$. In the superdiffusive framework, this propagator is non-Gaussian, and can be obtained in Fourier-Laplace space by an extension of the so-called Montroll-Weiss equation \citep{Metzler00,Metzler04,Zumofen93,Zimbardo13}. Indeed, the probability distribution of the jump lengths $\psi(\ell,t)$ in the Fourier-Laplace space can be written as
\be
\tilde{\psi}(k,s)=1-\tau s+C_1[(s+ikv)^{\mu}+(s-ikv)^{\mu}],
\ee
where $\tau$ is the average jump time, $k$ represents the wavenumber, and $s$ the Laplace transform variable. If we make an expansion to the lowest order both in $s$ and $k$, which implies the long time, long distance limit, and consider that $1 < \mu  < 2$, we obtain
\begin{equation}
\tilde{\psi}(k,s)\simeq 1-\tau s+C_2|k|^{\mu}
\label{eqpsi-fl2}
\end{equation}
(see \citet{Zimbardo13} for the constants $C_1$ and $C_2$).
Inserting this expression in the Montroll-Weiss equation and taking the inverse Laplace transform, one obtains the leading order term of the propagator for L\'evy walks in Fourier space \citep{Zimbardo13}, 
\be
\tilde{P}_{LW}(k,t)\sim {1\over2\pi} \exp{[-Ct|k|^{{\mu}}]},
\label{propk}
\ee
where $C$ is a constant (see below). 
The analytical inversion of equation (\ref{propk}) is only possible in two asymptotic cases. First, close to the source, that is for $x\ll(Ct)^{1/\mu}$, one has a modified Gaussian,
\be
P_{LW}(x,t) \simeq {\Gamma((\mu+1)/\mu) \over \pi (Ct)^{1/\mu}} \exp \left[-\frac{\Gamma(3/\mu)}{2\Gamma(1/\mu)} \left [{x\over (Ct)^{1/\mu}}\right ]^2 \right]
\label{eqpropgauss}
\ee
(see \citet{Zumofen93} and \citet{Zimbardo13} for a derivation).
Conversely, for large distances from the source, $x \gg (Ct)^{1/\mu}$ but $|x| < vt$, one obtains to leading order a power-law,
\be
P_{LW}(x,t) \simeq {\Gamma (\mu+1) \over \pi} \sin \left ({\pi\over2}\mu \right)  {C t \over |x|^{\mu+1}} ,
\label{eqpropxt}
\ee
(see also \citet{Zaslavsky02}).  In addition, for $|x| > vt$ the propagator smoothly goes to zero \citep{Blumen90}, this being the main difference with the L\'evy flight propagator given in the next section. 

With the additional, simplifying assumption as made by \citet{Zimbardo13}, that $\psi (\ell, t)$ can be approximated to be constant for $|\ell| \le \ell_0$, it can be shown that
\be
C = 2 {\mu - 1 \over \mu + 1} \left| \cos\left ({\pi\over2}\mu\right) \right| \Gamma(-\mu) {\ell_0^{\mu} \over t_0}.
\label{constant_C}
\ee
As shown in \citet{Zimbardo13}, the superdiffusion coefficient can be determined as
\be
{\cal D}_{\alpha}=\frac{2(2-\alpha)}{(3-\alpha)(4-\alpha)}\Gamma(\alpha-1) \ell_0^{2-\alpha} v^{\alpha}.
\label{Diff_alpha}
\ee
We can see that, basically, the value of ${\cal D}_{\alpha}$ depends on $\mu$ (i.e. ${\alpha}$) and on $\ell_0$. While $\mu$ alone determines the exponent of superdiffusion in Eq. (\ref{eq:alpha-mutil}), both $\mu$ and the scale parameter $\ell_0$ determine the values of the superdiffusion coefficient that sets the scale of transport. We estimate ${\cal D}_{\alpha}$ directly from spacecraft observations.

\section{Modelling based on a fractional diffusion-advection equation}

Although the jump length and jump duration coupling introduced in the
L\'evy walk approach above is in a physically strict sense more
appropriate, a model based on L\'evy flights, which requires no
coupling between the jump length and duration, can be
regarded as an effective model for this process, similar to the
diffusion approximation in classical energetic particle
transport. This type of model has the advantage that a fractional
Fokker-Planck type equation for the transport processes can be easily
formulated, which is analogous to transport equations employed in the
Gaussian diffusion context. This kind of equation can give solutions
for the complete space-time evolution of the particle distribution
function, which can be obtained by various complementary solution
techniques \citep{Litvinenko14}. In addition, for systems where the effective random walker velocity can be
different from the particle velocity, the decoupled L\'evy flight model is adequate
\citep{del-Castillo04,Perrone13,Bovet14}. This is the case, for instance, of transport perpendicular to the magnetic field due to the guiding centre motion caused by the $\mathbf{E}\times \mathbf{B}$ drift in laboratory plasmas: if the fluctuating electric field $\delta \mathbf{E}$ is strongly varying and bursty, as typical of turbulence, the guiding centre velocity will be strongly varying, too, and the coupling between $\ell$ and $t$ does not hold.

Here, we only state the general idea of the model and establish the
connection to the propagator of the L\'evy walk process as discussed
above. We refer to \citet{Magdziarz07}, \citet{Effenberger14},
\citet{Fichtner14}, \citet{Litvinenko14}, and \citet{Stern14}
for more details on the actual solutions and modelling with these equations.

A one-dimensional example of a fractional diffusion-advection
equation for shock-accelerated particles is given by
\begin{equation}
  \label{eq:diff-adv}
\frac{\partial}{\partial t} f(x,t) = V_a \frac{\partial}{\partial x} f(x,t) + \kappa \frac{\partial^{\mu}}{\partial |x|^{\mu}} f(x,t) + \delta(x) \,,
\end{equation}
which can be regarded as a simplified version of a more general
fractional Fokker-Planck equation \citep{Metzler00}. Here $f(x,t)$ is
the distribution function of energetic particles dependent on space
$x$ and time $t$, $V_a$ is a constant advection speed, and $\kappa$ is
the constant fractional diffusion coefficient with the 
dimension (length)$^\mu$/(time). In contrast to the case of normal diffusion, in the case of anomalous diffusion the constant $\kappa$, appearing in the fractional diffusion equation, and ${\cal D}_{\alpha}$, appearing in the expression of the mean square deviation, are not the same. These constants even have different physical dimensions. 
Particles are injected at the shock
with a delta-function source term. We introduced the fractional
Riesz derivative $\frac{\partial^{\mu}}{\partial |x|^{\mu}}$ with the
fractional index $\mu$ as a generalization of the one-dimensional
Laplace operator. It can be expressed by the (right and left hand)
Riemann-Liouville fractional derivatives, defined as 
\begin{equation}
  \label{eq:riemann-liouville-left}
  \tensor*[_{a}]{D}{_{x}^{\mu}} f(x) = \frac{1}{\Gamma(m-\mu)}\frac{\partial^m}{\partial x^m}\int_{{a}}^x\frac{f(x')}{(x-x')^{1+\mu-m}} dx' \,,
\end{equation}
and
\begin{equation}
  \label{eq:riemann-liouville-right}
  \tensor*[^{b}]{D}{_{x}^{\mu}} f(x) = \frac{(-1)^m}{\Gamma(m-\mu)}\frac{\partial^m}{\partial x^m}\int_{{x}}^b\frac{f(x')}{(x'-x)^{1+\mu-m}} dx' \,,
\end{equation}
with $m-1 < \mu < m$, where $m$ is an integer. The Riesz-derivative
\citep{Gorenflo99,Metzler00,Perrone13} is then given by
\begin{equation}
  \label{eq:riesz-riemann}
  \frac{\partial^{\mu}}{\partial |x|^{\mu}} f(x) = -\frac{1}{2\cos(\mu\pi /2)} (\tensor*[_{-\infty}]{D}{_x^{\mu}} + \tensor*[^\infty]{D}{_{x}^{\mu}}) f(x) \,.
\end{equation}
\subsection{The propagator for L\'evy flights}
A Green's function (or propagator) $P_{LF}(x,t)$ for the fractional
diffusion problem has to satisfy the fractional diffusion equation
with delta source. Neglecting the advection term in Equation (\ref{eq:diff-adv}), we get
\begin{equation}
  \label{eq:Greens-eq}
\frac{\partial}{\partial t} P_{LF}(x,t) = \kappa \frac{\partial^{\mu}}{\partial |x|^{\mu}} P_{LF}(x,t) + \delta(x)\delta(t) \,.
\end{equation}
Its Fourier transform is given by \citep[e.g.][]{Metzler00}
\begin{equation}
\label{eq:fourier-Greens-eq}
\partial_t\tilde{P}_{LF} = -\kappa |k|^\mu \tilde{P}_{LF} + \frac{1}{2\pi}\delta(t)  \,,
\end{equation}
so that we recover the Fourier form of the $\mu$-stable distribution as
\begin{equation}
\label{eq:fourier-Greens}
\tilde{P}_{LF}(k,t) = \frac{1}{2\pi} \exp(-\kappa|k|^\mu t) \,.
\end{equation}
%
%
%
After computing the inverse Fourier transform, an asymptotic expression for $P_{LF}(x,t)$ and $x\gg(\kappa t)^{1/\mu}$ is
given by \citep{Levy25,Zaslavsky02},
\begin{equation}
\label{eq:Greens-approx}
P_{LF}(x,t) \approx  \frac{\Gamma(\mu+1)}{\pi} \sin\left(\frac{\pi}{2}\mu\right)  \frac{\kappa t}{|x|^{\mu+1}},
\end{equation}
thus establishing the $-(1+\mu)$ asymptotic power-law behaviour of the
propagator. Comparing these equations with the approximate expressions for the L\'evy
walk propagator in Equation~(\ref{propk})--(\ref{eqpropxt}), we see that the
fractional diffusion equation can be regarded as an approximate
model for L\'evy walks by identifying $C$ as $\kappa$ and having the same value of 
${\mu}$. Therefore, the fractional diffusion coefficient $\kappa$ in Equation (\ref{eq:Greens-eq}) basically depends on $\ell_0$ and $\mu$ (see Equation (\ref{constant_C})).
The propagators for L\'evy flights and L\'evy walks differ because the tails of the latter go to zero for $|x| > vt$.
\subsection{The fractional superdiffusion coefficient}
Given the analogy found for the two propagators in Equations (\ref{eqpropxt}) and (\ref{eq:Greens-approx}),
a relation between the ``anomalous diffusion
coefficient'' ${\cal D}_\alpha$ (see Equation~\ref{Diff_alpha}) and $\kappa$
can be derived. A simple dimensional analysis would give $({\cal
  D}_\alpha)^{1/\alpha} \propto \kappa$ with $\alpha=2/\mu$ (see also
the discussion in \citet{Metzler00}). However, we can recover Equation~\ref{eq:alpha-mutil} from the calculation of the mean square displacement considering a finite particle speed $v$.
Although the formal mean square displacement diverges for
L\'evy flights, one can derive a constrained or ``moving'' mean square
displacement from the asymptotic expression (\ref{eq:Greens-approx})
for the Green's function. Consider the definite integral
\begin{eqnarray}
\label{eq:moving-mean}
\langle x^2\rangle_\mathrm{m} &=& \int_{-vt}^{vt}x^2 P_{LF}(x,t) dx \nonumber\\
&&\approx \frac{2}{\pi} \sin\left(\frac{\pi}{2}\mu\right) \Gamma(1+\mu) \kappa t \int_{0}^{vt}|x|^{1-\mu}  dx \nonumber\\
&=& \frac{2}{\pi (2-\mu)} \sin\left(\frac{\pi}{2}\mu\right) \Gamma(1+\mu) \kappa |v|^{2-\mu}  t^{3-\mu} \,.
\end{eqnarray}
Upon comparing with the general expression for the
mean square displacement, we obtain the relation for the superdiffusion exponent of L\'evy
walks, $\alpha = 3-\mu$. Furthermore, a relation between
the diffusion constant ${\cal D}_\alpha$ and the fractional diffusion
coefficient $\kappa$ follows as 
\begin{eqnarray}
\label{eq:D-kappa}
{\cal D}_\alpha = \frac{1}{\pi (\alpha-1)} \sin\left(\frac{\pi}{2}(3-\alpha)\right) \Gamma(4-\alpha) |v|^{\alpha-1} \kappa \,.
\end{eqnarray}
Using Eqs. (\ref{constant_C}) and (\ref{Diff_alpha}) and the reflection formula for the Euler gamma function, one can verify that the above coefficients are sufficiently close so that the L\'evy flight scenario can be used to describe the L\'evy walks to a good approximation. Thus, besides the superdiffusion coefficient, we can also derive $\kappa$ from data analysis, once having computed ${\cal D}_\alpha$ from Equation (\ref{Diff_alpha}). This determination of $\kappa$ is useful for further applications of the fractional equation formalism to energetic particle transport. 
The relation between ${\cal D}_\alpha$ and $\kappa$ given here is different from that shown in Equation (35) in \citet{Metzler98} or in Equation (86) in \citet{Metzler00}, which only holds for subdiffusion.

\begin{table*}
\centering
\caption{Examples of heliospheric shocks which exhibit upstream power-law intensities of energetic particle fluxes}
\begin{tabular}{cccccc}
\hline
Event & s/c & Date & heliocentric  & energetic & energies \\ 
  \#  &     &      & distance [AU] & particles &  [MeV]   \\ 
\hline
 1 &  Ulysses   & 11/10/1992$^a$ &  5.2 & electrons & 0.04--0.29  \\
 2 &  Ulysses   & 22/01/1993$^c$ &  5.0 & electrons & 0.06--0.18  \\
 3 &  Ulysses   & 10/05/1993$^c$ &  4.7 & electrons & 0.04--0.29  \\
 4 &  Voyager~2 & 30/08/2007$^d$ & 83.7 & ions      & 0.54--3.50  \\
\hline
\end{tabular}

\footnotesize{$a:$ \citet{Perri09a}; $b:$ \citet{Perri08};\\ $c:$ \citet{Perri07}; $d:$ \citet{Perri09b}.}
\label{table1}
\end{table*}

\begin{table*}
\centering
\caption{Measured parameters of the heliospheric shocks listed in Table \ref{table1}}
\begin{tabular}{ccccccc}
\hline
\# & $V_1^{\rm s/c}$ & $V_2^{\rm s/c}$ & $n_1$  & $n_2$ & $\theta_{\rm Bn}$ $\!\!\!^a$ & $t_{\rm break}^{\rm s/c}$\\ 
   &      [km/s]     &      [km/s]     & [cm$^{-3}$] & [cm$^{-3}$] &                   & [hours] \\ 
\hline
1 & $755\pm 10$ & $643\pm 4$ & $0.13681\pm 0.00007$   & $0.3617\pm 0.0006$   & 68$^{\circ} \pm 11^{\circ}$ &   8 \\
2 & $760\pm 1$ & $583\pm 4$ & $0.110236\pm 0.000004$   & $0.572\pm 0.006$   & 35$^{\circ} \pm 13^{\circ}$ &  20 \\
3 & $814\pm 50$ & $645\pm 311$ & $0.1171\pm 0.0003$   & $0.352\pm 0.002$   & 59$^{\circ} \pm 10^{\circ}$ &   4 \\
$4\;^b$ & $300$ & $155$ & $0.001$ & $0.00158$ & 74$^{\circ}$--82$^{\circ}$  & 240 \\
$4\;^c$ & $368$ & $145\pm 112$ & $9.6194\pm 0.0006 (10^{-4})$ & $2.0757\pm 0.0002 (10^{-3})$ & 74$^{\circ}$--82$^{\circ}$  & 240 \\
\hline
\end{tabular}

\footnotesize{For events~1--3: data from CDAWeb; for event~4: data from space.mit.edu/pub/plasma/vgr/v2/daily;\\
 $a:$ from \citet{Balogh95}. $b$: plasma values from \citet{Richardson08};\\
 $c$: plasma values computed over a sliding window of $7$ days over a $100$ days period.\\
 For event $4$, when the statistical error is $>100\%$ only the error values are reported. }
\label{table2}
\end{table*}

\section{Application to spacecraft measurements}
We now directly estimate the superdiffusion coefficient ${\cal D}_\alpha$ and the fractional diffusion coefficient $\kappa$, derived from the two approaches described in Sections 2 and 3, from spacecraft observations for a number of interplanetary shock waves. We selected those events for which particle superdiffusion has already been observed \citep{Perri08,Perri09a,Perri09b,Zimbardo12}.

\subsection{The scale parameter}
Recall that the paramater $\ell_0$ indicates the scale at which the jump probability $\psi(\ell,t)$ turns out to have a power-law shape.
In the observations, this is related to the distance from the shock at which the particle time profiles
begin to exhibit a power-law behaviour, too. There is indeed a break between the shape of the particle time profile in the vicinity of the
shock and its shape at few hours upstream from the shock front (see Figure 1) and this corresponds to the two asymptotic cases for the propagator which have been shown in
Eqs. (\ref{eqpropgauss}) and (\ref{eqpropxt}).
\begin{figure}[h!]
\begin{center}
\includegraphics[width=0.7\columnwidth]{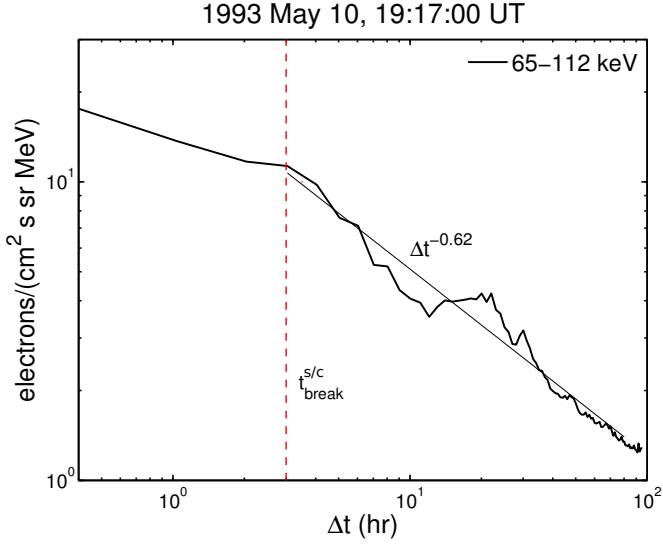}%
\end{center}
\caption{Hourly resolution electron flux measured by the Ulysses spacecraft upstream of the CIR shock crossing of May 10, 1993 (solid line). The time $\Delta t$ represents the difference between the upstream measurement time and the shock crossing time. The power-law fit is represented by the thin solid line, and the time of the break in the power-law by the dashed red line.
\label{f1}}
\end{figure} 

Considering the form of $C$ in Eq.(\ref{constant_C}) we can define a scaling variable $\xi = (x/\ell_0)/(t/t_0)^{1/\mu}$ \citep{Zumofen93,Zimbardo13}, where $t_0=\ell_0/v$, so that we have $\xi \ll 1$ close to the shock. The modified Gaussian propagator in Eq. (\ref{eqpropgauss}) can be used, giving a nearly flat profile (indeed, for $\xi \ll 1$ the Gaussian is slowly varying). Conversely, $\xi \gg 1$ represents a region farther away from the shock and the power-law form of the propagator in Eq. (\ref{eqpropxt}) can be used, giving a power-law particle profile as described by \citet{Perri07,Perri08}. Clearly, the change from one form to the other corresponds to the case $\xi \approx 1$. 
Therefore, it is possible to obtain the distance $x_{\rm break}$ of the break, which separates the different forms of the upstream energetic particle profiles from the condition $\xi=1$. Thus, we find
\be
\frac{x_{\rm break}}{\ell_0}=\biggr(\frac{t_{\rm break}}{t_0}\biggl)^{1/{\mu}}.
\label{eqxi2}
\ee
Further, the relation $\ell_0 = v t_0$ is implied from the space-time coupling present in Eq. (\ref{eq:psi}), so that giving
the scale length $\ell_0$ is equivalent, for particles of given velocity known from spacecraft measurements, to giving the scale time $t_0$. Since we are deriving the parameters for energetic particles accelerated at interplanetary shocks, we denote the upstream and downstream values by the subscripts 1 and 2,
respectively, and the reference frame i.e. spacecraft or shock, by the superscripts `s/c' or `sh', respectively.
Assuming a steady-state shock, the particle motion upstream of the shock can be described as a competition between the supersonic plasma advective motion, $x=V_1^{\rm sh} t$, and the superdiffusive random motion $\langle \Delta x^2 \rangle = 2 {\cal D}_{\alpha} t^{\alpha}$. Then, we insert in Eq. (\ref{eqxi2}) an expression for $x_{\rm break}$ that comes from the advective motion, namely
\be
x_{\rm break} = V_1^{\rm sh} t_{\rm break}.
\label{eqbreakbreak}
\ee
Using Eq. (\ref{eqbreakbreak}) and $\ell_0 = v t_0$ and rewriting in terms of distances, we obtain from Eq. (\ref{eqxi2})
\be
x_{\rm break} = \ell_0 \left ({v\over V_1^{\rm sh}} \right )^{1/(2-\alpha)}
\label{break}
\ee
which gives the distance of the break from the shock (see Figure \ref{f1}). 
Equation (\ref{break}) compares with the expression for the exponentiation length 
$L_{\rm prec}= D/V_1^{\rm sh} = {1\over3}\lambda v / V_1^{\rm sh}$, which is obtained in the case of normal diffusion for a constant diffusion coefficient \citep[e.g.][]{Lee82}. However, it is worth remembering that $\ell_0$ does not have the same meaning as the mean free path $\lambda$, which diverges in the superdiffusive case.
The higher the upstream plasma flow velocity
 $V_1^{\rm sh}$, the shorter either $L_{\rm prec}$ or $x_{\rm break}$. This is because the advective motion is
`squeezing' energetic particles towards the shock. We may notice that studying the precursor, i.e.\ the profile of
upstream particles, is the best and most direct way to study the transport properties perpendicular to the shock
(provided the upstream $V_1^{\rm sh}$ is known).

\subsection{The energy spectral index of particles accelerated at the shocks}
The transport of energetic particles also influences the shape of the energy spectrum. In particular, DSA predicts an energetic particle spectrum that only depends on the compression ratio of the shock \citep{Drury83}. This holds for both relativistic and non-relativistic particles. Thus, the investigation of the energy spectra for particles accelerated at shock waves can be a test for the validity of DSA. Recent observations during the TS crossings by the two {\it Voyager} spacecraft gave indications of a compression ratio at the shock of $r\sim 2$, although large-scale temporal variations of the TS makes it difficult to accurately estimate the value of the compression ratio. \citet{Decker08} have shown that the slope of the differential spectrum over a broad range of energy channels is $\sim 1.26$. This value can be recovered by DSA, assuming $r\sim 3$, which is in strong disagreement with the value obtained from {\it Voyager} observations. Recently, \citet{Arthur13} numerically integrated the focussed transport equation for a spherical, stationary shock that includes both a precursor and a subshock. They assumed a global compression ratio of about $2.8$ . The energy spectra obtained were in good agreement with those derived from  the {\it Voyager} data, thus indicating that particles with diffusion length larger than the whole shock size (including the precursor and the subshock) will experience a compression ratio greater than $2$. This result is consistent with a DSA scenario, however the effective compression ratio sampled by particles can be energy dependent \citep{Amato14} so that low energy particles will experience a compression ratio closer to that observed at the subshock i.e. $r\sim 2$.

We propose an alternative explanation based on superdiffusion for the spectral index values determined for particles accelerated at the TS, and for differential energy spectra of particles accelerated at a couple of CIRs reported in Table \ref{table1} (see details below).
\citet{Perri12a} derived the particle spectral index in the framework of the superdiffusive shock acceleration (SSA):  in this case the spectral index depends on both the compression ratio and the index of superdiffusion $\alpha$. For ultra-relativistic particles the differential energy spectral index is
\be
\gamma=\frac{6}{r-1}\frac{2-\alpha}{3-\alpha}+1,
\label{gammaUR}
\ee
while for non-relativistic particles \citet{Perri12a} obtained
\be
\gamma=\frac{3}{r-1}\frac{2-\alpha}{3-\alpha}+1.
\label{gammaNR}
\ee
\citet{Perri12a} further calculated the spectral index $\gamma_j=\gamma-0.5$ of the differential flux $dj/dE$ for non-relativistic particles assuming superdiffusive transport; with a compression ratio $r=2$ and $\alpha=1.3$, the value found in \citet{Perri09b} for the particles accelerated upstream of the TS--see Table \ref{table3}, we obtained a spectral index of $\sim 1.7$ (see Table \ref{table4}). This value is very close to the $1.67$ value found in the low energy charged particle spectrum detected by {\it Voyager 1} just after the TS crossing \citep{Decker05}, but not in a very good agreement with the $1.26$ value reported from the {\it Voyager 2} observations \citep{Decker08}. Thus, an increased compression ratio, as considered by \citet{Arthur13}, can bring the predictions closer to the observations. Still, SSA is able to give harder spectral indices so that $r=2.4$ would be sufficient to explain the observed $\gamma_j$, assuming $\alpha=1.3$.
Below we will compare the spectral indices predicted by SSA with the indices directly obtained from the observed energy spectra of particles accelerated at CIRs.
Electrons accelerated at CIR events in Table \ref{table1} are relativistic, so that the measured differential flux (assuming that the particle speed within each energy channel is almost constant) can be written as
\be
\frac{dj}{dE}=v\frac{dN}{dE}\equiv v\frac{dN}{dp}\frac{dp}{dE}\propto vp^{-\gamma_p}\frac{dp}{dE},
\label{eqJ}
\ee
where $\gamma_p$ is the index of the differential spectrum in momentum. After manipulating Eq. (\ref{eqJ}), being $E=\sqrt{p^2c^2+m^2c^4}$, it can be easily derived 
\begin{eqnarray}
\lefteqn{dj/dE\sim p^{-\gamma_p}} \nonumber \\
\lefteqn{\gamma_p=[(6/(r-1))(2-\alpha)/(3-\alpha)]+1.}
\label{eqdjde}
\end{eqnarray}
Since for the events in Table \ref{table1} both $r$ and $\alpha$ are known, $\gamma_p$ can be computed and compared with the slope of the $dj/dE$ spectra observed.

\subsection{Data analysis and method}

For interplanetary shocks measurements are made by spacecraft as a function of time, so that we need to estimate
$x_{\rm break}$ by the observed break time $t_{\rm break}^{\rm s/c}$ as seen by the spacecraft. This is not the break time for particle advection introduced in Eq. (\ref{eqbreakbreak}), but the break in the particle spatial profile as seen by the spacecraft when sampling different distances from the shock, as shown in Figure \ref{f1}. Thus, the break position is obtained as 
\be
x_{\rm break} = V_{\rm sh}^{\rm s/c}  t_{\rm break}^{\rm s/c}
\ee
where $V_{\rm sh}^{\rm s/c}$ is the relative velocity between the spacecraft and the shock. 

To estimate the upstream $V_1^{\rm sh}$ in the shock frame we have to take the following transformation of velocities into account:
\be
V_1^{\rm s/c} = V_{\rm sh}^{\rm s/c} + V_1^{\rm sh}.
\ee
We make the simplifying assumption that the shocks are planar and perpendicular to the radial direction, even though it is well known that CIR shocks are oblique with respect to the radial direction \citep[e.g.][]{Gosling99}. A study of the full set of Rankine-Hugoniot jump conditions is deferred to future work. 
Considering further that the
compression ratio is to be expressed in the shock frame as $r= V_1^{\rm sh}/V_2^{\rm sh}$, we can easily obtain
the relation \citep[e.g.][]{Burgess95}
\be
V_{\rm sh}^{\rm s/c} = {r V_2^{\rm s/c} - V_1^{\rm s/c}\over r - 1 } \, .
\ee
The compression ratio can be obtained from the observed densities as $r=n_2/n_1$, although strong fluctuations of the measured parameters, like plasma density, velocity, etc.\, are usually found (see the discussion in \citet{Balogh95} and in \citet{Giacalone12}). For example, in the hourly averaged Ulysses data, the plasma density can change by 50\% in a few hours. Therefore, to estimate of the plasma parameters upstream and downstream of the shock waves, we calculated both plasma speed and density within a time window of seven hours for those CIRs detected by Ulysses for which superdiffusion of energetic electrons was observed. These Ulysses events are listed in Table \ref{table1}. After a first iteration, the window is moved by one hr, then the calculation of velocity and density is repeated. This process continues by spanning a $12$-$24$ hours time interval upstream and downstream (that is shifting the time window for $12$-$24$ hours), depending on the ``regularity'' of the velocity and density profiles in each event. To avoid sharp variations of the plasma parameters close to the shock fronts, we started the above evaluation procedure after a couple of hours from the time of the shock, both upstream and downstream. Finally, we calculated an average value for the plasma speed and the density from the values computed within the running windows, along with a standard deviation (statistical error); these are reported in Table \ref{table2}. Although the procedure described here for the plasma parameters calculation is accurate, the intrinsic strong variability of the time series upstream and downstream of the shocks can lead to large unavoidable errors in their estimation.
When the standard deviations are larger than the averages, only the error values are given. 

For the TS crossing (event~4 in Table \ref{table1}), the analysis required more attention. For a first estimation, we considered the shock and plasma parameters given by \citet{Richardson08} at the so-called TS-3 event (i.e. shock speed in the spacecraft frame $V_{\rm sh}^{\rm s/c}\sim -68$ km/s, and compression ratio of the shock $r\sim 1.58$). In particular, we estimate velocities and densities upstream and downstream of the TS from the profiles in Figure 3 in \citet{Richardson08} in the vicinity of the shock crossing. For a second estimation, we also considered a larger scale trend of the plasma parameters to be consistent with the analysis of energetic proton time profiles in \citet{Perri09b}. Indeed, a power-law decay for the proton time profiles was found for about 100 days  upstream of the TS. Thus, we have further calculated the plasma densities and speeds over sliding windows of seven days for a $100$ days period upstream and downstream of the TS crossing on 2007 September 1st. In this case, we assigned to the shock a speed (in the spacecraft frame) of $-27$~km/s. This double estimation is because, while for TS-3 Voyager~2 crossed the inwards moving TS, which was a 
typical quasi-perpendicular shock with shock normal angle $\theta_{Bn}= 74.3^{\circ}$ \citep{Richardson08}, other TS crossings exhibited different parameters. Therefore, to estimate an average shock speed, we also considered the results of a global MHD simulation of the TS dynamics, performed by \citet{Washimi07}: considering the solar wind parameters of 2006 and 2007, they found that the shock is moving
towards the sun (inwards). Thus, the relative velocity between the spacecraft and the shock is given in the reference frame of the spacecraft by $V_{\rm sh}^{\rm s/c}=|V_{\rm sh}|+V_{\rm s/c}\sim 27$ km/s, being
$V_{\rm sh}\sim -12$ km/s the TS's inwards speed with respect to the Sun \citep{Washimi07}, and $V_{\rm s/c}\sim 15$ km/s the spacecraft's speed with respect to the Sun. From Table \ref{table3}, it can be seen that both the break distance $x_{\rm break}$ and the scale length $\ell_0$ depend on the assumed shock speed.

For each particle species, i.e. electrons for events~1--3 and protons for event~4, we can obtain the speed from the energy
\be
v = c \sqrt{1-{m^2 c^4\over E^2}}
\ee
where $E=mc^2 + E_k$ is the total free particle energy, with $E_k$ the `kinetic' energy determined by the spacecraft instruments. All energies can be conveniently expressed in keV.

An expression for the scale length $\ell_0$ can now be obtained by inverting the relation (\ref{break}) above,
\be
\ell_0 = x_{\rm break} \left ({V_1^{\rm sh}\over v } \right )^{1/(2-\alpha)} .
\label{eq_ell_0}
\ee
The value of $\alpha=3-\mu$ is obtained by the slope of the power-law upstream time profile for each energy channel \citep{Perri07,Perri08,Perri09a,Perri09b}. To estimate the particle speed, we used the average energy of each energy channel. Once the value of $\ell_0$ is known, the values of ${\cal D}_{\alpha}$ and $\kappa$ can be obtained.

\begin{figure}[h!]
\begin{center}
\includegraphics[width=0.7\columnwidth]{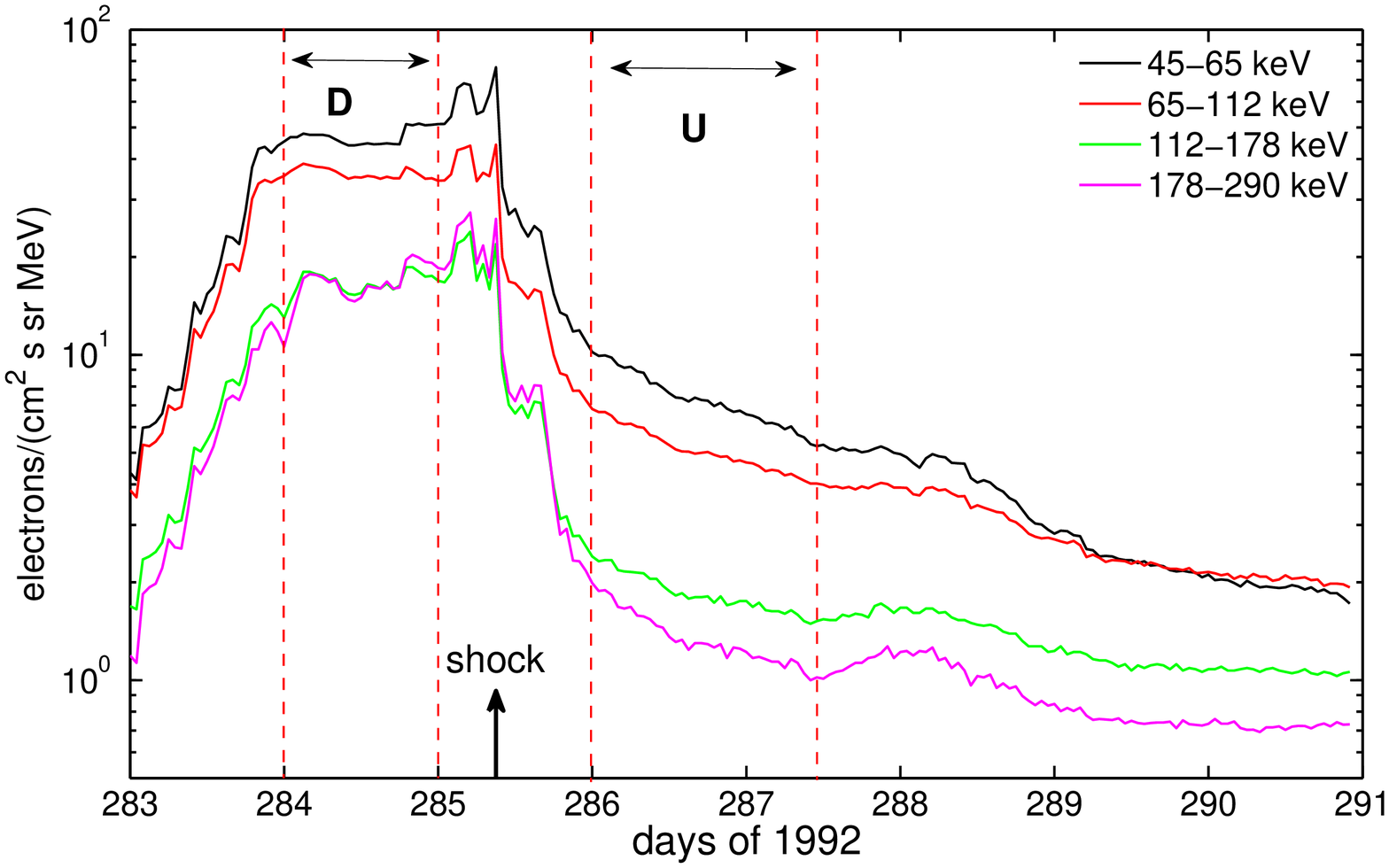}%
\end{center}
\caption{Hourly resolution electron flux measured by the Ulysses spacecraft at the CIR shock of 11/10/1992. The shock position is indicated by the vertical thick arrow. The vertical red dashed lines delimit the regions upstream and downstream where an average flux $<dj/dE>$ has been computed.
\label{f2}}
\end{figure}

\begin{figure}[h!]
\begin{center}
\includegraphics[width=0.7\columnwidth]{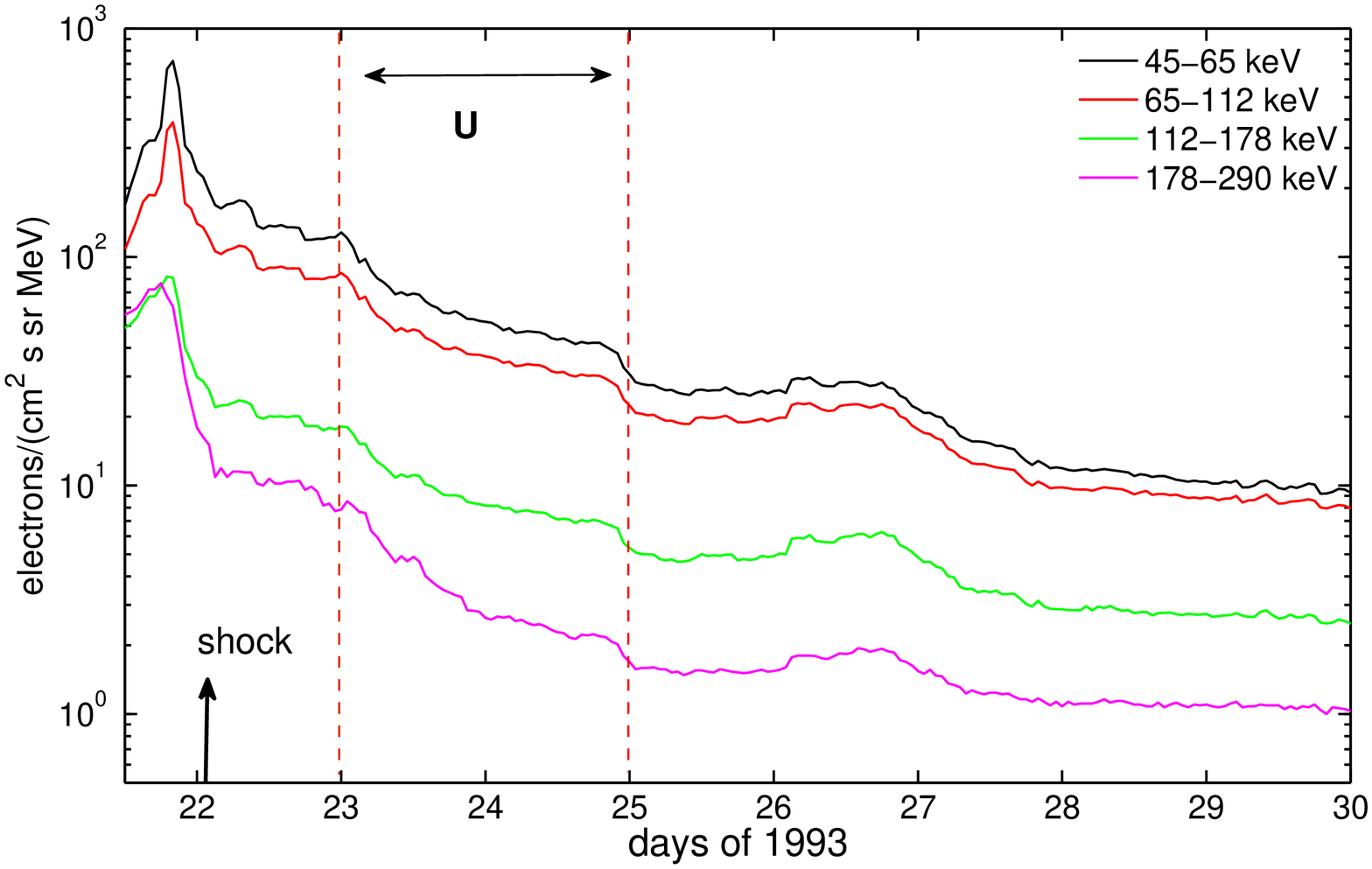}%
\end{center}
\caption{Same as Figure \ref{f2}, except for the CIR shock of 22/01/1993.
\label{f2b}}
\end{figure}

We further computed the particle energy spectral indices for the CIR events $1$ (both upstream and downstream) and $2$ (upstream). The downstream fluxes of event $2$ and the event $3$ in Table \ref{table1} have not been analyzed for obtaining $\gamma_p$ (see eqs.\ref{eqJ}, \ref{eqdjde}) since the electron time profiles fall down rapidly, thus making unreliable the computation of an average flux over a time window of few hours. Moreover, the data within the different energy channels tend to be almost overlapped, making it impossible to obtain a clear scaling in the energy spectrum. The reasons for the high variability and the overlapping of the particle time profiles are not clear and require further investigation.
For electron time profiles of event $1$ (Figure \ref{f2}) and for upstream time profiles of event $2$ (Figure \ref{f2b}), we calculated an average differential flux $<dj/dE>$ over the time windows delimited by the vertical red dashed lines in Figures \ref{f2} and \ref{f2b} (the shock positions are indicated by the thick vertical arrows). 
Figure \ref{f3} shows the obtained $<dj/dE>$ as a function of the ``momentum'' $cp=\sqrt{E_k^2+2mc^2E_k}$ for event $1$. The kinetic energy $E_k$ corresponds to the `centre' value of each energy channel. The same plot is shown in Figure \ref{f4}, but for event $2$. The slopes of the differential energy spectra are reported in Table \ref{table4} along with the theoretical values expected by DSA and SSA once the compression ratio of the shock and the exponent $\alpha$ of superdiffusion are known.

\subsection{Results of data analysis}
%
\begin{table*}[h!]
\centering
\caption{Derived parameters for the heliospheric shocks listed in Table \ref{table1}}
\begin{tabular}{p{0.2cm}p{0.7cm}p{0.7cm}p{0.7cm}p{1.0cm}cccccc}
\hline
\# & $V_{\rm sh}^{\rm s/c}$ & $V_1^{\rm sh}$ & $V_2^{\rm sh}$ & $x_{\rm break}$ & $E_k$ & $v$ & $\alpha$ & $\ell_0$ & ${\cal D}_{\alpha}$ & $\kappa$\\ 
   &                        & $\!\!\![$km/s]         &                & $\!\!\![10^6$km]     & [keV] & [$^a$] & & [km] & [$^b$] & [$^c$] \\ 
\hline
1  & 576 & 179 &   68 & 16.6 & 53.5 & 127 & $1.44\pm 0.08$ &   620 &  450  & 4 \\
   &     &     &      &      & 88.5 & 157 & $1.56\pm 0.09$ &  30 &  220 & 0.5  \\
   &     &     &      &      &  145 & 188 & $1.7\pm 0.1$ & 0.03 &  80 & 0.04  \\
   &     &     &      &      &  234 & 218 & $1.6\pm 0.2$ & 7 &  300 & 0.3  \\
\\
2  & 541 & 219 &   42 & 38.9 & 53.5 & 127 & 1.00 & -- &   --  & -- \\
   &     &     &      &      & 88.5 & 157 & $1.08\pm 0.02$ & 15000 & 20000 & 10000 \\
   &     &     &      &      &  145 & 188 & $1.19\pm 0.03$ &  9000 & 5000  & 600 \\
   &     &     &      &      &  234 & 218 & $1.02\pm 0.05$ & 30000 & 130000 & 100000 \\
\\
3  & 560 & 219 &   73 &  8 & 53.5 & 127 & $1.29\pm 0.08$ &   9000 &  2000  & 100 \\
   &     &     &      &      & 88.5 & 157 & $1.38\pm 0.07$ &   2000 &  1000  & 20 \\
   &     &     &      &      &  145 & 188 & $1.31\pm 0.08$ &   4000 &  2000  & 100 \\
   &     &     &      &      &  234 & 218 & $1.15\pm 0.08$ &  15000 &  10000 & 2000 \\

\\
4  & -68 & 368 &  233 &   59 &  765 &  12 & $1.30\pm 0.07$ & 400000& 1500 & 20 \\
   &     &     &      &      & 1565 &  17 & $1.29\pm 0.08$ & 270000& 2000 & 25 \\
   &     &     &      &      & 2820 &  23 & $1.3\pm 0.2$ & 100000& 1500 & 15 \\
   & -27 & 396 &  183 & 23   &  765 &  12 & $1.30\pm 0.07$ & 180000&  3500 & 300 \\
   &     &     &      &      & 1565 &  17 & $1.29\pm 0.08$ &  100000&  4500 & 350 \\
   &     &     &      &      & 2820 &  23 & $1.3\pm 0.2$ &  60000&  2500 & 150 \\
\hline
\end{tabular}

\footnotesize{Notes on units: $a:$ $10^3$ km/s; $b:$ $10^6$ km$^2$/s$^{\alpha}$; $c:$ $10^6$ km$^{\mu}$/s. Errors have also been estimated for derived quantities (not shown).\\ For $\ell_0$, ${\cal D}_{\alpha}$, and $\kappa$ the statistical errors are larger than their mean values, therefore only the error values are reported. 
              }
\label{table3}
\end{table*}

The results of our analysis are summarized in Table \ref{table2} (measured parameters with errors) and Table \ref{table3} (derived parameters with errors) for both the interplanetary shock crossings by Ulysses (events~1--3) and for the TS crossing by Voyager 2 (event~4).

Table \ref{table3} shows that for event~1 the obtained values of $\ell_0$ range from 30 m to 600 km, depending on the energy channel. Although the particle velocity enters the expression of $\ell_0$, most of this variability is due to the variability of $\alpha$, which determines the exponent in Eq. (\ref{eq_ell_0}), and to the small value of the ratio $V_1^{\rm sh}/v\sim 10^{-3}$.
The values of $\ell_0$ reported are actually those associated with statistical errors, since errors are always larger than $100\%$. The analysis shown in \citet{Perri09a} clearly indicates that the energetic particle time profiles upstream of the shock decay as power laws, rather than as simple exponential (as envisaged by DSA), thus superdiffusion is expected. However, the exponent of the mean square displacement $\alpha$, which gives the degree of superdiffusion, can vary according to the particle energy. Indeed, different energies imply different Larmor radii and then a different (resonant) interaction between particles and magnetic fluctuations. In turn, this implies the large variability of $\ell_0$.
The values of the diffusion coefficients ${\cal D}_{\alpha}$ and $\kappa$ (computed using the $\ell_0$ values in Table \ref{table3}) change in the same ``direction'' as $\ell_0$, but with much smaller deviation. In particular, the values of ${\cal D}_{\alpha}$ are rather stable, a property that is important for developing predictive algorithms for the fluxes of solar energetic particles.

A similar behaviour is found for events~2 and~3. On the other hand, event~2 is only slightly superdiffusive ($\alpha = 1.0$--$1.19$) and this gives larger values of $\ell_0$ and of the diffusion coefficients. In the limit of normal diffusion, ${\cal D}_{\alpha}$ and $\kappa$ cannot be computed because $\Gamma(\alpha-1)$ diverges. While this divergence hints at making the above coefficients closer to the large diffusion coefficient usually obtained for normal diffusion, one has to remember that most of the expressions given in Section 2 are valid for $\alpha$ strictly larger than one. Also, event~2 is only marginally superdiffusive. This can be explained by considering that this event is a quasi-parallel shock (see Table \ref{table2}-- $\theta_{\rm Bn}\sim 35^{\circ}$). Thus, an upstream ion foreshock is formed, which enhances the level of turbulent fluctuations and wave activity due to the backstreaming reflected ions. This results in a stronger interaction between particles and fluctuations so that there is a higher probability of recovering a normal diffusive scenario \citep{Pommois07}. 

In the case of the TS event, $\alpha$ shows similar values in all the energy channels; in this case $\ell_0$, ${\cal D}_{\alpha}$, and $\kappa$ do not broadly vary but tend to be more stable.

Despite the large variability of $\ell_0$, we note that $x_{\rm break}$, which is directly determined from the spacecraft data, does not exhibit such a large variability but is quite stable from shock to shock.
It is important to remark that $\ell_0$ is the particle jump length above which the probability distribution of the particle jump lengths i.e. $\psi(\ell,t)$, decays as a power-law (see Eq.(3)). For each event, Table \ref{table3} shows that higher values of the exponent $\alpha$ of the mean square displacement correspond  to smaller values of $\ell_0$. When the value of $\alpha$ increases (but always bounded $1<\alpha<2$ in superdiffusive processes) the particle propagation becomes more and more superdiffusive, therefore $\ell_0$ might become shorter and shorter, so that the particle during its propagation has a higher probability of making both very long and very short jumps. This gives rise to a very non-homogeneous, scale-free propagation of particles in space. 
%
\begin{figure}[h!]
\begin{center}
\begin{minipage}{\columnwidth}
\includegraphics[width=0.8\columnwidth]{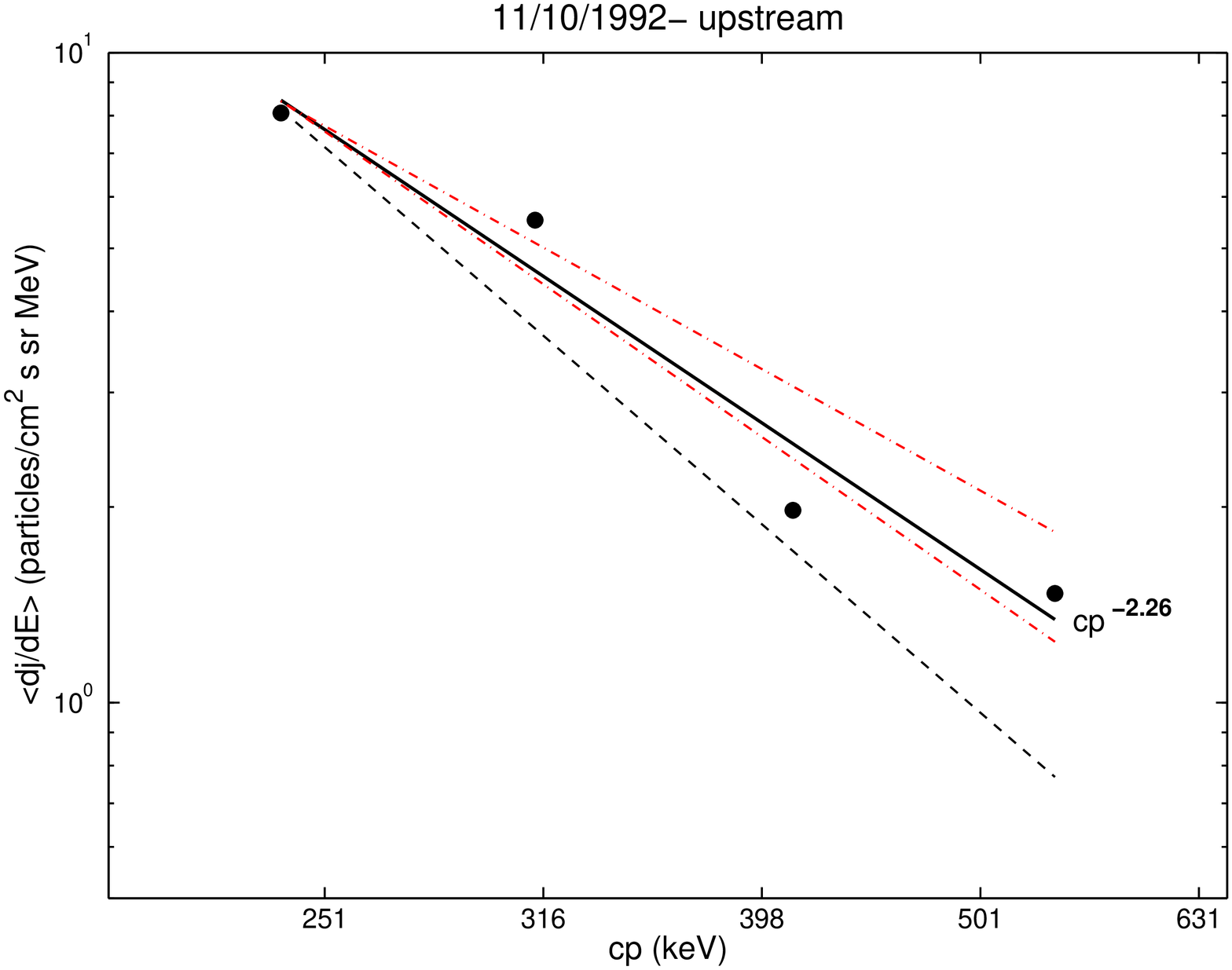}%
\end{minipage}
\begin{minipage}{\columnwidth}
\includegraphics[width=0.8\columnwidth]{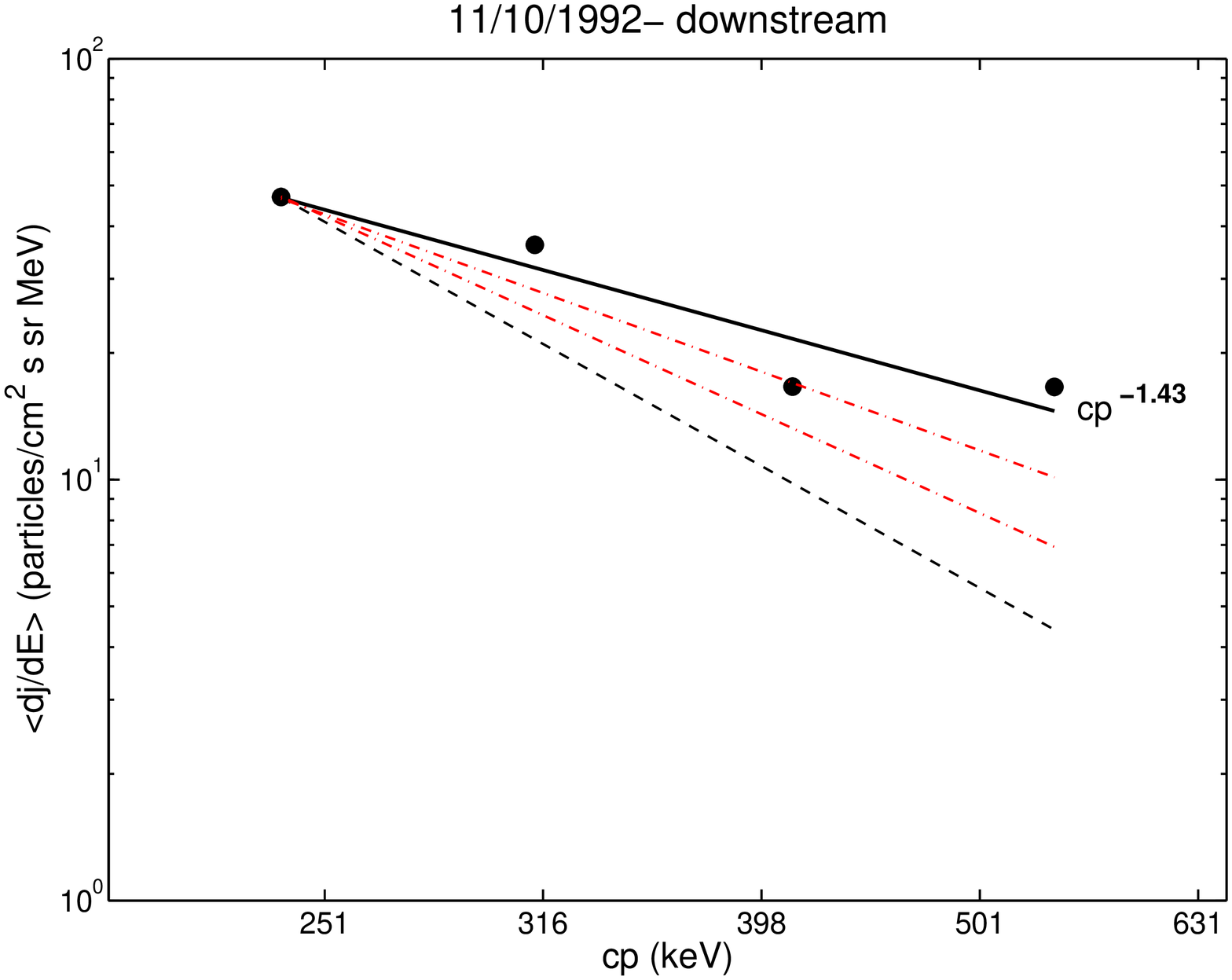}%
\end{minipage}
\end{center}
\caption{Energy spectra (filled circles) for the CIR shock of 11/10/1992 calculated upstream (top panel) and downstream (bottom panel). The thick solid line represents the best power-law fit. The black dashed lines indicate the DSA prediction and the red dash-dotted lines indicate the power-law behaviour predicted by SSA considering the minimum and the maximum values of $\alpha$ found from the electron time profiles (see text).
\label{f3}}
\end{figure} 
\begin{figure}[h!]
\begin{center}
\includegraphics[width=0.7\columnwidth]{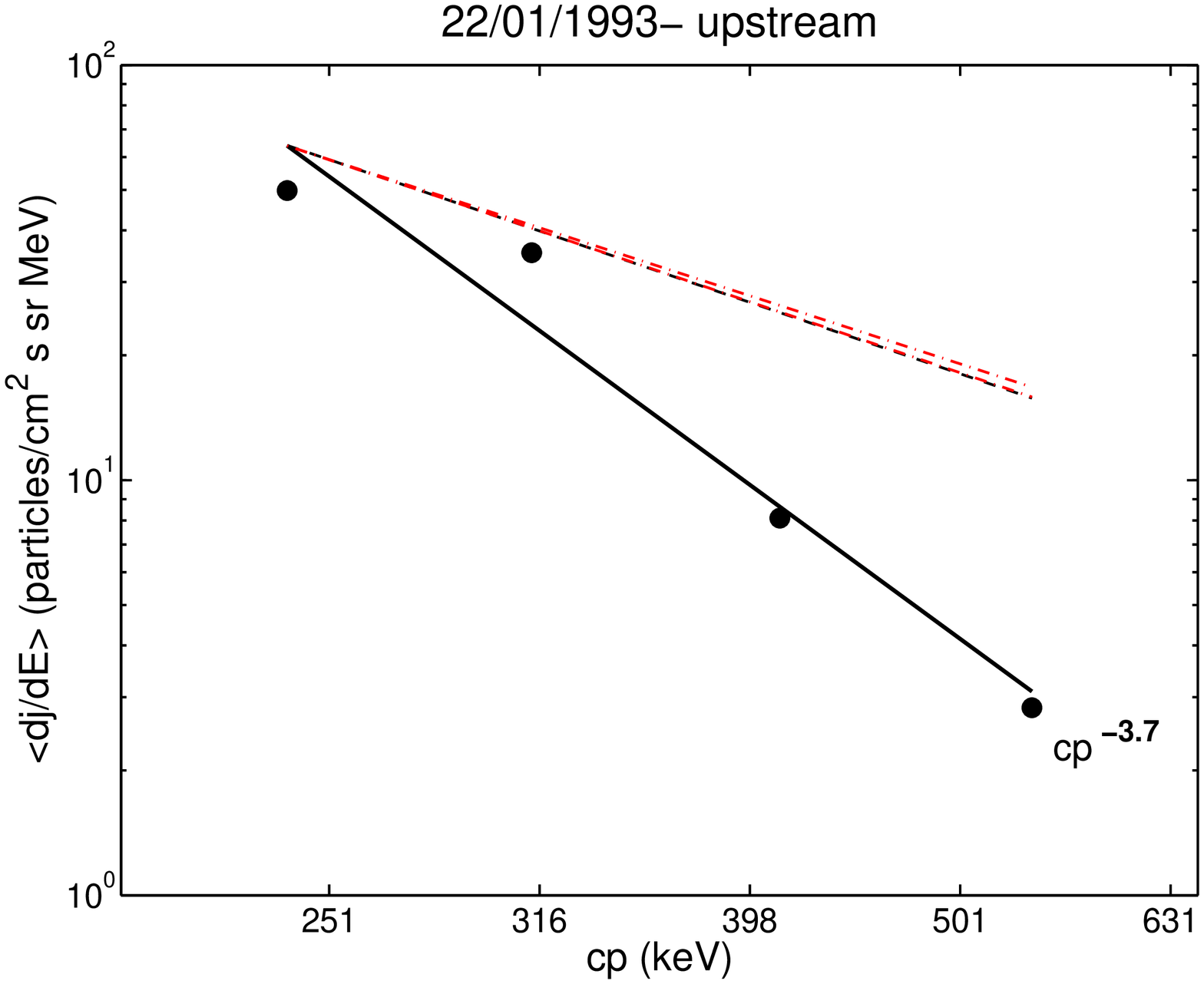}%
\end{center}
\caption{Same as Figure \ref{f3}, except for the CIR shock of 22/01/1993.
\label{f4}}
\end{figure} 

The electron energy spectra for the CIR of 11/10/1992 are shown in Figure \ref{f3}; in particular, the top panel in Figure \ref{f3} displays the energy spectrum (filled circles) derived from $<dj/dE>$ upstream of the shock, while the bottom panel reports the spectrum computed downstream. The best power-law fits are indicated by the thick solid lines (the slopes are also displayed in the panels). The black dashed lines indicate the trend expected from DSA, while the red dotted-dashed lines refer to the power-law behaviour predicted by SSA considering the minimum and the maximum $\alpha$ value found by the analysis of the electron time profiles (see Table \ref{table3}). It can be noted that the observed energy spectra, both upstream and downstream, are well described by the SSA model, although the agreement is much better upstream. This can be explained by the fact that we have calculated $\gamma_p$ downstream using the values of $\alpha$ found for particle time profiles upstream. 
The same plot has been produced for the CIR event of 22/01/1993 (see Figure \ref{f4}). However, in this case neither the DSA and the SSA are able to reproduce the observed electron energy spectrum, which results to be much steeper. Notice that for this event the compression ratio is $>4$ (see Table \ref{table4}), suggesting non-stationarity and a large time variability for the plasma parameters upstream and downstream of the shock.
As discussed above, a comparison between the {\it Voyager} observations reported by \citet{Decker08} and the values predicted by the theoretical models gives indication of a better agreement with the SSA model (as indicated in Table \ref{table4}).

\begin{table*}[t!]
\centering
\caption{Results from the analysis of the electron spectra.}
\begin{tabular}{cccccc}
\hline
\# & r & $\alpha$ & $\gamma_{\rm obs}$ & $\gamma_{\rm DSA}$ & $\gamma_{\rm SSA}$\\
\hline
1 U & 2.6 & [1.44-1.70] & 2.26 & 2.9 & [1.87-2.35]\\
1 D & 2.6 & [1.44-1.70] & 1.43 & 2.9 & [1.87-2.35]\\
2 U & 5.19 & [1.02-1.19] & 3.7 & 1.72 & [1.64-1.71]\\
4 & 2.0 & 1.3 & 1.26 & 2.0 & 1.7 \\
\hline
\end{tabular}      
\label{table4}
\end{table*}

\section{Conclusions}
In conclusion, we have presented the description of anomalous, superdiffusive transport in terms of a probabilistic description, corresponding to L\'evy walks, and in terms of a fractional transport equation, corresponding to L\'evy flights. Although they represent two different approaches to anomalous transport, we have shown similarities between them and also how the mean square deviation foreseen for L\'evy walks can be recovered from the solution of the fractional transport equation. These results are interesting because the use of the fractional transport equations has extensively been applied to a variety of physical systems, including laboratory plasmas \citep{del-Castillo04,Bovet14,Litvinenko14}. Further, it is important to understand to what extent L\'evy flights can be used to recover the properties of L\'evy walks. Within those frameworks, parameters of superdiffusive transport, as the corresponding diffusion coefficients ${\cal D}_{\alpha}$ and $\kappa$, have been expressed in terms of the scale parameter $\ell_0$. The latter and, in turn, the superdiffusion  and the fractional diffusion coefficients have, for the first time, been estimated directly from spacecraft observations at heliospheric shocks. 
The values of $\ell_0$ reported in Table \ref{table3} are found to be much shorter than the mean free path $\lambda$ which is, according to the standard estimates \citep[e.g.][]{Bieber94}, of the order of $0.1$--$1$ AU at a heliographic distance of 1 AU, and of the order of a few tens of AU at the TS \citep{Giacalone13}. These low values of the scale length $\ell_0$ have been interpreted as indicative of very short acceleration times (see Eq. (40) in \citet{Zimbardo13}).
It is however important to recall that $\ell_0$ does not have a meaning similar to the mean free path $\lambda$: firstly, the mean free path is diverging in the case of superdiffusion, and secondly, $\ell_0$ is the shortest length for free paths having a power-law distribution. This means that free paths much longer than $\ell_0$ have relative high probability, and their occurrence gives rise to superdiffusion. 
We have also checked the expression of the spectral index predicted by SSA for the considered shocks. Within the experimental uncertainties, SSA gives a much better explanation for the observed $\gamma_{\rm obs}$ values than DSA, thanks to the capability to give harder spectral indices than DSA.
The results obtained give insight into the random walk properties for superdiffusion, furnish an estimation of the input parameters for numerical simulations, and supply the basis for further studies of energetic particle acceleration at heliospheric and astrophysical shocks.

\begin{acknowledgements}
We highly appreciate discussions at the team meeting `Superdiffusive transport in space plasmas and its influence on energetic particle
acceleration and propagation' supported by the International Space Science Institute (ISSI) in Bern, Switzerland, and with Yuri Litvinenko.
S.P.'s research has been supported by ``Borsa Post-doc POR Calabria FSE 20 07/2013 Asse IV Capitale Umano - Obiettivo Operativo M.2.'' H.F.\ acknowledges support by the German Research Foundation (DFG) via the project FI~706/14-1. Authors thank the referee for the constructive criticism.
\end{acknowledgements}

\end{document}